# ARTICLE

## Multiscale Modeling of Gas Adsorption and Surface Coverage in Thermocatalytic Systems

*Jikai Sun,[a] and Jianzhong Wu*[a]*

Conventional methods for modeling thermocatalytic systems are typically based on the Kohn-Sham density functional theory (KS-DFT), neglecting the inhomogeneous distributions of gas molecules in the reactive environment. However, industrial reactions often take place at high temperature and pressure, where the local densities of gas molecules near the catalyst surface can reach hundreds of times their bulk values. To assess the environmental impacts on surface composition and reaction kinetics, we integrate KS-DFT calculations for predicting surface bonding energy with classical DFT to evaluate gas distribution and the grand potential of the entire reactive system. This multiscale approach accounts for both bond formation and non-bonded interactions of gas molecules with the catalyst surface and reveals that the surface composition is determined not only by chemisorption but also by the accessibility of surface sites and their interactions with the surrounding molecules in the gas phase. This theoretical procedure was employed to establish the relationship between surface coverage, gas-phase composition, and bulk phase thermodynamic conditions with thermocatalytic hydrogenation of $CO_2$ as a benchmark. The computational framework opens new avenues for studying adsorption and coverage on catalytic surfaces under industrially relevant conditions.

## Introduction

Adsorption of reactant species — whether by physisorption or chemisorption — on catalyst surfaces is a prerequisite for heterogeneous catalytic reactions.[1,2] Whereas the adsorption isotherms are often described using statistical-mechanical models that account for the surface energy and intermolecular interactions, quantum chemistry calculations are essential for understanding bond formation underlying chemisorption and thermocatalytic reactions. The reaction mechanisms are governed by both chemisorption and non-bonded interactions, which depend on the properties of the active sites, the energy barriers to reaction, and the local concentrations of reactants. Recently, an increasing number of studies have emphasized the significant impact of the reaction environment and co-adsorbed species on the surface coverage.[3,4] For instance, KS-DFT calculations revealed the strong influence of co-adsorbed HCOO* on Cu catalysts for $CO_2$ hydrogenation to methanol.[5] Likewise, it has been demonstrated that oxygen coverage on Pd plays a crucial role in determining the kinetics of hydrogen oxidation.[6] While the importance of gas adsorption and surface coverage in thermocatalytic reactions has been well documented, a comprehensive framework for describing bond formation and various forms of intermolecular forces dictating thermocatalytic processes has yet to be developed.

Existing methods for calculating surface coverage can be broadly categorized into two theoretical approaches. The first approach employs kinetic Monte Carlo (KMC) simulations based on the energy profiles of elementary reactions predicted by KS-DFT.[7] The stochastic method describes catalytic processes based on the transition rates of adsorption, desorption, surface diffusion, and elementary reactions.[8] The temperature and pressure effects are typically considered through the Arrhenius equation for kinetic processes and the Langmuir isotherm for chemisorption.[9] While in principle KMC can model surface interactions at a microscopic level, it often assumes that the interactions between co-adsorbed species can be described in terms of surface coverage or site-blocking. The second approach relies on KS-DFT calculations for the surface energies of various possible adsorption states to identify the most thermodynamically stable surface coverage.[10] Given the vast number of surface configurations, exhaustive consideration of every scenario with KS-DFT calculations is impractical. As a result, simplified models are often employed, such as the two-line model proposed by Hu and coworkers.[11,12]

For both approaches discussed above, an accurate description of bond formation is a prerequisite for understanding surface coverage and reaction kinetics. In KS-DFT calculations, the chemisorption energy is typically obtained from the change of the ground-state energies due to bond formation between the adsorbate and the catalyst surface:

$$E_{ad} = E_{*\text{adsorbate}} - E_* - E_{\text{adsorbate}} \qquad (1)$$

where $E_*$, $E_{\text{adsorbate}}$, and $E_{*\text{adsorbate}}$, represent the ground-state energies of a pristine surface, the adsorbate in the vacuum, and the surface bonded with the adsorbate, respectively. The first principles method neglects the influence of other molecular species in the

[a.] J. Sun, J. Wu
Department of Chemical and Environmental Engineering, University of California, Riverside, CA 92521, USA
E-mail: jwu@engr.ucr.edu

reaction environment and thus does not account for co-adsorption effects or the physical adsorption of gas molecules. However, as discussed below, due to the strong attraction between the catalyst surface and surrounding gas molecules, the surface concentration of gas-phase species can reach several hundred times of the bulk value. Neglecting the environmental effect may lead to a significant pressure gap.[13] For example in modeling thermocatalytic conversation of $CO_2$, the conventional approach often faces with difficulties in addressing issues such as the variation of adsorption sites with temperature and pressure,[14] the inhibitory effect of CO coverage on catalytic activity,[15,16] and the behavior of CO coverage under ultra-high pressure conditions.[17,18]

In our previous study based on KS-DFT calculations[19], we found that an isolated $H_2O$ molecule shows strong adsorption onto a $SiO_2$ surface in an aqueous environment, contradicting the well-known fact as the surface is hydrophobic. By introducing a water-box model and performing *ab initio* molecular dynamics (AIMD) simulations, we discovered that water molecules initially adsorbed on the $SiO_2$ surface were pulled back into the liquid phase due to hydrogen bonding with surrounding water molecules. The drastic difference in the adsorption energy indicates that the attraction of water molecules to the $SiO_2$ surface is much weaker than their affinity for the bulk liquid, confirming the hydrophobic nature of $SiO_2$. These findings highlight the critical role of the liquid-phase environment, which should be explicitly considered in KS-DFT calculations of the adsorption energy. In recent years, this issue has garnered increasing attention, leading to a growing number of AIMD studies that incorporate explicit liquid-phase environments using water-box models.[20,21]

For thermocatalytic reactions in gas-phase systems, it is commonly assumed that gas molecules exhibit low density and that intermolecular interactions are insignificant, allowing their influence to be neglected in predicting the adsorption energy. Consequently, KS-DFT calculations for gas-phase reactions are typically conducted without considering non-bonded intermolecular interactions.[22] While this assumption is justified at low pressure, it fails to account for the strong accumulation of gas molecules to the catalyst surface. In our previous work[23], we found that the surface concentration of $CO_2$ on Cu surfaces can be several hundred times higher than its gas-phase bulk concentration. At industrially relevant conditions, the concentrations of gas-phase molecules at the catalyst surface are highly inhomogeneous and deviate significantly from the bulk condition. Because the interactions between gas-phase molecules on the catalyst surface have important impacts on both the adsorption behavior and reaction kinetics, modeling gas-phase adsorption and surface coverage must account for not only bond formation but also the non-bonded interactions of the adsorbate with gas molecules in the bulk phase. Furthermore, the catalyst surface is generally covered by gas-phase molecules due to physical adsorption, and chemisorption causes the redistribution of these pre-existing gas molecules. The significant changes in free energy associated with the redistribution of gas molecules must be accounted for in understanding chemisorption and reaction kinetics.

In this work, we integrate the KS-DFT and classical DFT (cDFT) simulations to establish the relationship between surface coverage, gas-phase composition, and bulk phase thermodynamic conditions. This multiscale procedure accounts for both bond formation and non-bonded interactions during gas adsorption and catalytic reactions, including the presence of gas-phase components occupying the catalyst surface. Using thermocatalytic hydrogenation of $CO_2$ on several catalysts as a benchmark, we analyze gas adsorption energies and surface coverages of various species across a wide range of thermodynamic conditions. The theoretical study provides insights into the variation in adsorption energies among different chemical species, the selection of preferred adsorption sites, and the mechanisms governing surface coverage and poisoning.

## Results and Discussion

The multiscale procedure that integrates KS-DFT with cDFT calculations has been reported in our previous study[23]. Briefly, KS-DFT predicts the bond formation energies, while cDFT accounts for

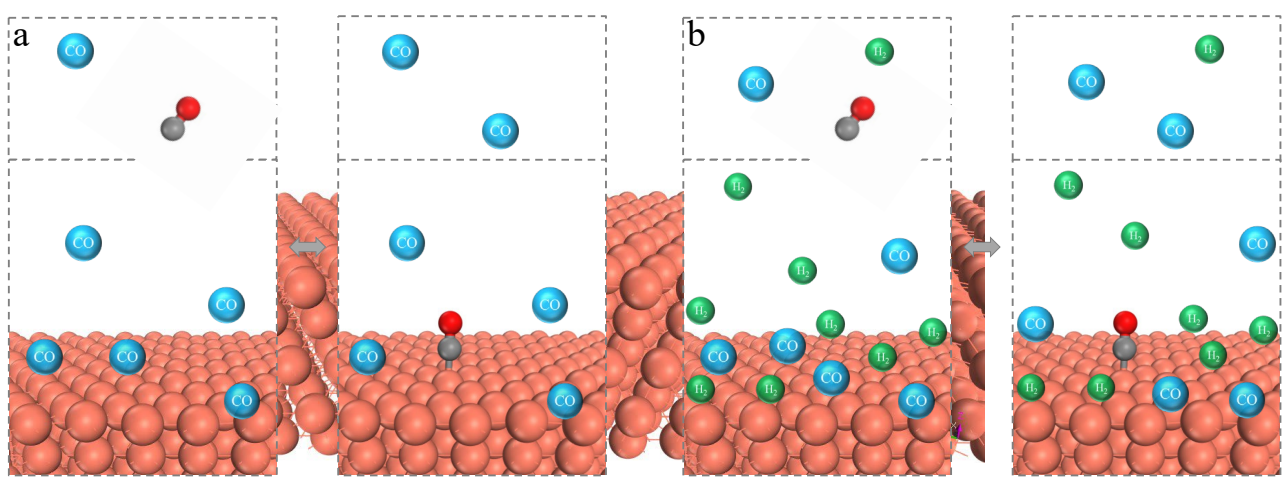

**Fig. 1** Scheme of the multiscale approach for modeling CO adsorption on a metal surface. (a) In a gas environment of pure CO, (b) In a gas mixture of $H_2$ and CO. The ball-and-stick model depicts a CO molecule bonded with the metal surface (orange ball), simulated using KS-DFT. Green and blue spheres indicate $H_2$ and CO molecules, respectively, modeled via cDFT. Note that the positions of the colored spheres are only illustrative; they represent a statistical distribution as described by statistical mechanics, rather than fixed atomic positions.

**Journal Name** ARTICLE

the interactions between gas molecules and catalyst surface through a semi-empirical force field. The environmental effects lead to a modified surface adsorption energy that depends on both the bond potential and the grand potential of the surrounding gas molecules

$$\Omega_{ad} = G_{ad} + \Omega_{cDFT-ad} \qquad (2)$$

where $G_{ad}$ is the adsorption free energy obtained from KS-DFT, and $\Omega_{cDFT-ad}$ represents the change in the grand potential of the gas phase due to the chemisorption. The latter is calculated via cDFT based on the Lennard-Jones model of gas molecules. The detailed computational procedure is described in the Supporting Information (SI).

In this work, $\Omega_{ad}$ is referred to as the adsorption grand potential, and $\Omega_{cDFT-ad}$ is called the penalty grand potential. While $G_{ad}$ incorporates the bonding energy $E_{ad}$ and the entropy correction of bond vibrations, $\Omega_{cDFT-ad}$ accounts for the free energy penalty for the displacement of gas-phase species on the surface during the chemisorption process.

It is important to note that both gas adsorption and surface coverage are sensitive to temperature, pressure, and gas-phase bulk composition. In traditional KS-DFT calculations, the pressure and composition effects are corrected using the equations for an ideal-gas mixture. Specifically, the pressure correction is implemented by subtracting $RT \ln P_i$ in the adsorption energy, where $R$ denotes the gas constant and $P_i$ is the partial pressure of adsorbate $i$. Meanwhile, the temperature effect is considered through the change in the bond vibrational entropy. Using the grand canonical ensemble for open systems, cDFT calculations account for interactions between different gas-phase components, the catalyst surface, and surface-adsorbed species under varying temperature, pressure and bulk composition, thereby reducing the pressure gap in modeling heterogeneous catalysis.

We may elucidate the physical significance of the adsorption grand potential ($\Omega_{ad}$) using CO adsorption on a metal surface as an example. As shown in Fig. 1a, $\Omega_{ad}$ represents the difference between the grand potential of the reactive system under two scenarios corresponding to CO transition from physisorption to chemisorption at the copper surface. In other words, $\Omega_{ad}$ corresponds to the change in the free energy due to the chemisorption of CO in the presence of other gas molecules. While $G_{ad}$ accounts for the chemisorption energy, $\Omega_{cDFT-ad}$ reflects its influence in the grand potential of the entire system. In Fig. 1b, we illustrate the chemisorption of a CO molecule from a gas mixture. Again, $\Omega_{cDFT-ad}$ arises from the "solvation" of the surface site due to its interactions with the surrounding molecules. It reflects the difference in no-bonded interaction energy between the gas-phase components and the adsorption site before and after CO bonding. This term is typically repulsive because the chemisorption often reduces the attraction between the metal site and the surrounding gas molecules. In this context, $\Omega_{ad}$ includes contributions from changes in both the bonding energy and the non-bonded interactions resulting from the redistribution of gas-phase molecules.

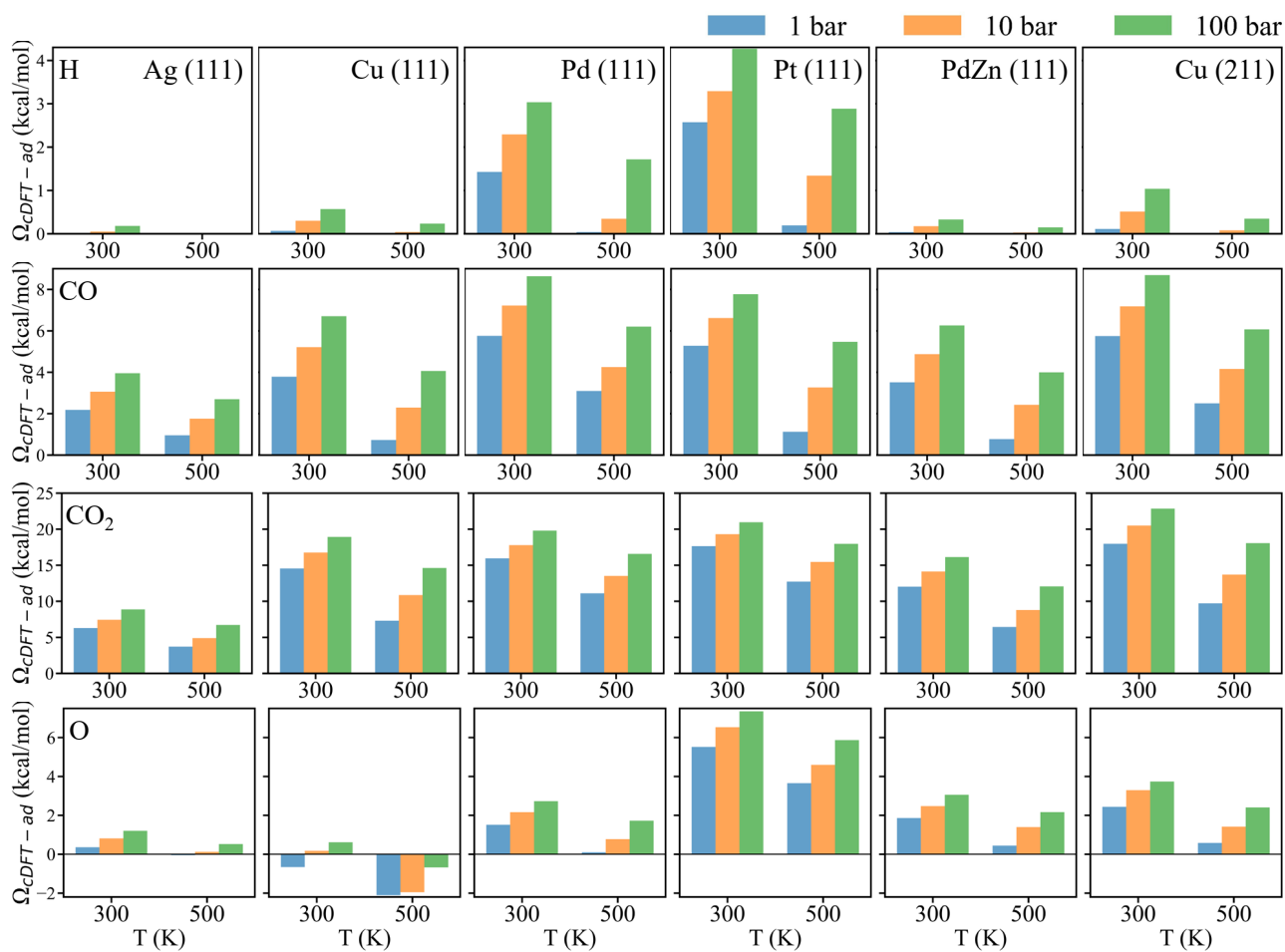

**Fig. 2** The penalty grand potential ($\Omega_{cDFT-ad}$) of various chemical species on different catalyst surfaces at varying temperature and pressure. The rows correspond to H, CO, $CO_2$, and O, respectively. The columns represent Ag (111), Cu (111), Pd (111), Pt (111), PdZn (111), and Cu (211), respectively.

*Environmental effects on the penalty grand potential*

We begin by examining how the reaction environment affects the penalty grand potential for the adsorption of different chemical species, using $CO_2$ hydrogenation as a benchmark. In this thermocatalytic process, H, CO, $CO_2$, and O are commonly considered as the main chemical species adsorbed on various catalyst surfaces.[10,24] As discussed above, the environmental effects on the adsorption energy is reflected in the penalty grand potential ($\Omega_{cDFT-ad}$), which is determined from cDFT calculations on KS-DFT structures.

Fig. 2 presents the penalty grand potential for the chemisorption of H, CO, $CO_2$, and O on six catalyst surfaces: Ag (111), Cu (111), Pd (111), Pt (111), PdZn (111), and Cu (211).[25–27] For all adsorbates and catalyst surfaces, $\Omega_{cDFT-ad}$ increases with rising pressure and decreasing temperature. The environmental correction follows the trend: Ag (111) < Cu (111) < Pd (111) < Pt (111), which aligns with the adsorption strength of these metals.[28,29] After Zn doping, the $\Omega_{cDFT-ad}$ value for PdZn (111) is smaller than that for Pd (111) for all adsorbates, consistent with the trend for the adsorption strength. In comparison between Cu (111) and Cu (211) surfaces, the latter exhibits a stepped surface configuration, leading to a larger penalty grand potential due to stronger interactions of gas molecules with Cu atoms. The enhanced penalty grand potential highlights the unique nature of the stepped surface, where the binding sites facilitate more extensive interactions with gas-phase molecules. This suggest that the high catalytic activity of the stepped surface stems not only from its under-coordinated atomic sites, but also from its enhanced attraction toward gas molecules.[9,30,31]

For all six catalyst surfaces considered in this work, $\Omega_{cDFT-ad}$ follows the same order: H < O < CO < $CO_2$. This trend can be explained in terms of the excluded volumes of these adsorbed species: The larger the adsorbed species, the greater the surface area it occupies, resulting in a larger penalty grand potential. Interestingly, for the oxygen atom, a negative value of $\Omega_{cDFT-ad}$ is observed on the Cu (111) surface, indicating that, after the surface site is occupied by an oxygen atom, more gas-phase molecules are adsorbed on the metal surface. This anomaly can be explained with the density profiles shown in Fig. 3. With the surface site occupied by an oxygen atom, the gas molecules are distributed closer to catalyst surfaces, leading to an increased surface concentration and adsorption energy. A similar behavior is observed for $H_2$ adsorption on the Ag (111) surface, as shown in Fig. 3a, indicating that the adsorption of smaller species such as H and O enhances the accumulation of gas-phase components on the catalyst surface. The different signs of the penalty grand potential could be further explained by the differential gas-phase densities illustrated on Fig. 3e-f. Although the local gas-phase density is reduced near the adsorbate, the density may increase in the region surrounding the adsorption site due to the edge effects. For the chemisorption of CO and $CO_2$, the adsorbate occupies a large surface area, reducing the net absorption of gas molecules at the catalyst surface. While for the chemisorption an oxygen atom, its excluded volume is relatively small. In comparison to other adsorbates, the environmental effects on H adsorption are rather insignificant not only because of the low surface density of $H_2$ and small adsorption energy, but also because the H adsorbate has minimal effects on the distribution of gas molecules near the catalyst surface.

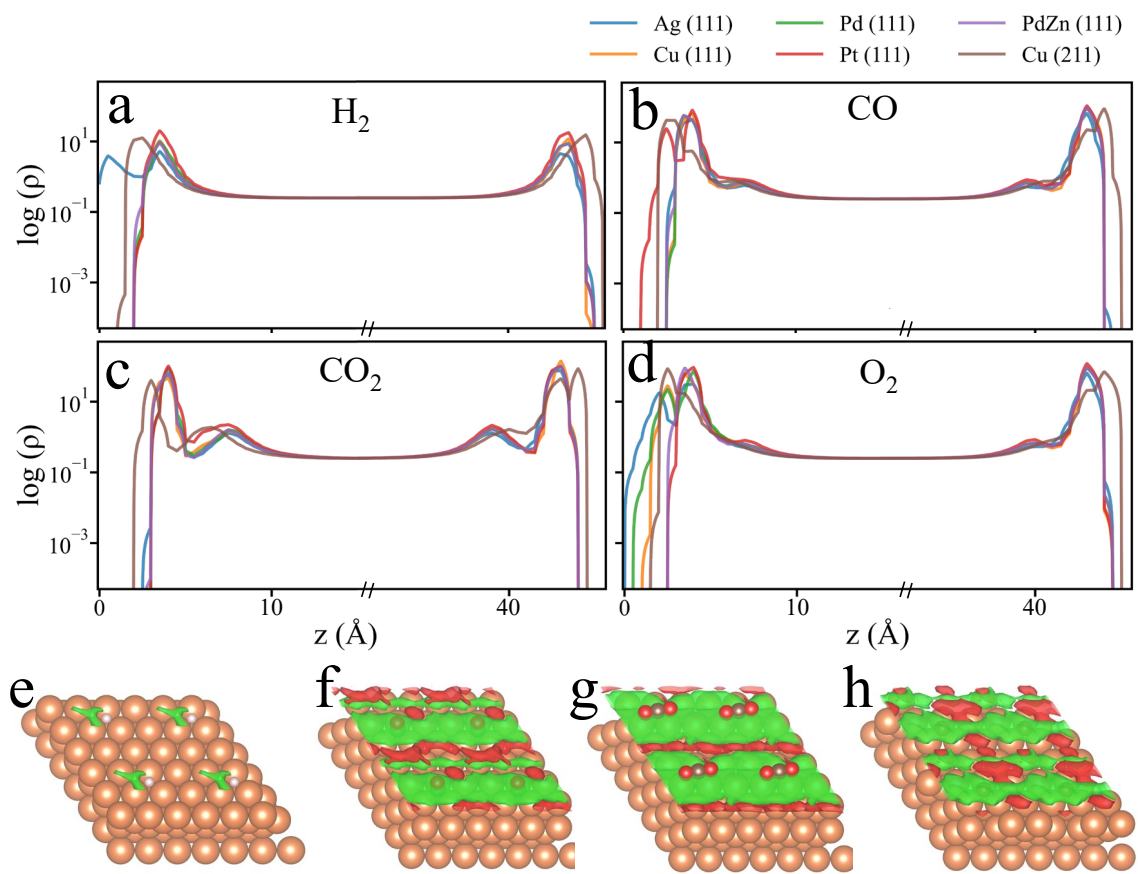

**Fig. 3** Density profiles of gas molecules near adsorbate (left) and pristine (right) catalyst surfaces. (a) chemisorption of H from $H_2$ gas, (b) chemisorption of CO from CO gas, (c) chemisorption of $CO_2$ from $CO_2$ gas, and (d) chemisorption of O from $O_2$ gas. For all cases, the gas temperature is 500 K and pressure is 10 bar. The z-axis starts from the top layer of the adsorbed surface ($z=0$ Å) and ends at the top layer of the pristine surface ($z=45$ Å). The differential gas-phase density due to the chemisorption of H (e), CO (f), $CO_2$ (g), and O (h) on Cu (111) surface. The differential gas-phase density is calculated from $\rho_{diff} = \rho_{*species} - \rho_*$, where $\rho_{*species}$ and $\rho_*$ are the gas-phase density on species adsorbed surface and pristine surface, respectively. The density iso-surface is set as 1 e⁻⁶ mole/Å³. Green color indicates the reduction of the gas density, and red indicates gas-phase density increase.

The above discussion suggests that the presence of small adsorbates on the surface can be beneficial to catalytic activity rather than detrimental, contrary to conventional assumptions of surface poisoning.[32–34] Because the penalty grand potential $\Omega_{cDFT-ad}$ is influenced by multiple factors, including the surface density of gas-phase components, the inherent adsorption strength of the catalyst surface, its surface morphology, and the size of the adsorbed molecules, the environmental effects on chemisorption and surface coverage can often be counterintuitive, challenging the predictions of conventional models.

*Preferential adsorption of CO on different surface sites*

The reaction environment can influence not only the adsorption energy but also bond formation at different surface sites, which affect the catalytic behavior. In the following analysis, using CO as a case study, we investigate the effect of surface-accumulated gas-phase molecules on CO adsorption at various binding sites on PdZn (111) surface.

As illustrated in Fig. 4, CO can chemically bind to the catalyst surface at four possible adsorption sites: top, bridge (bri), fcc, and hcp. Fig. 4a shows that for CO adsorption at the top site of the PdZn (111) surface, the penalty grand potential ($\Omega_{cDFT-ad}$) increases with increasing CO partial pressure and decreasing temperature. The trend is consistent with the surface density profile presented in Fig.

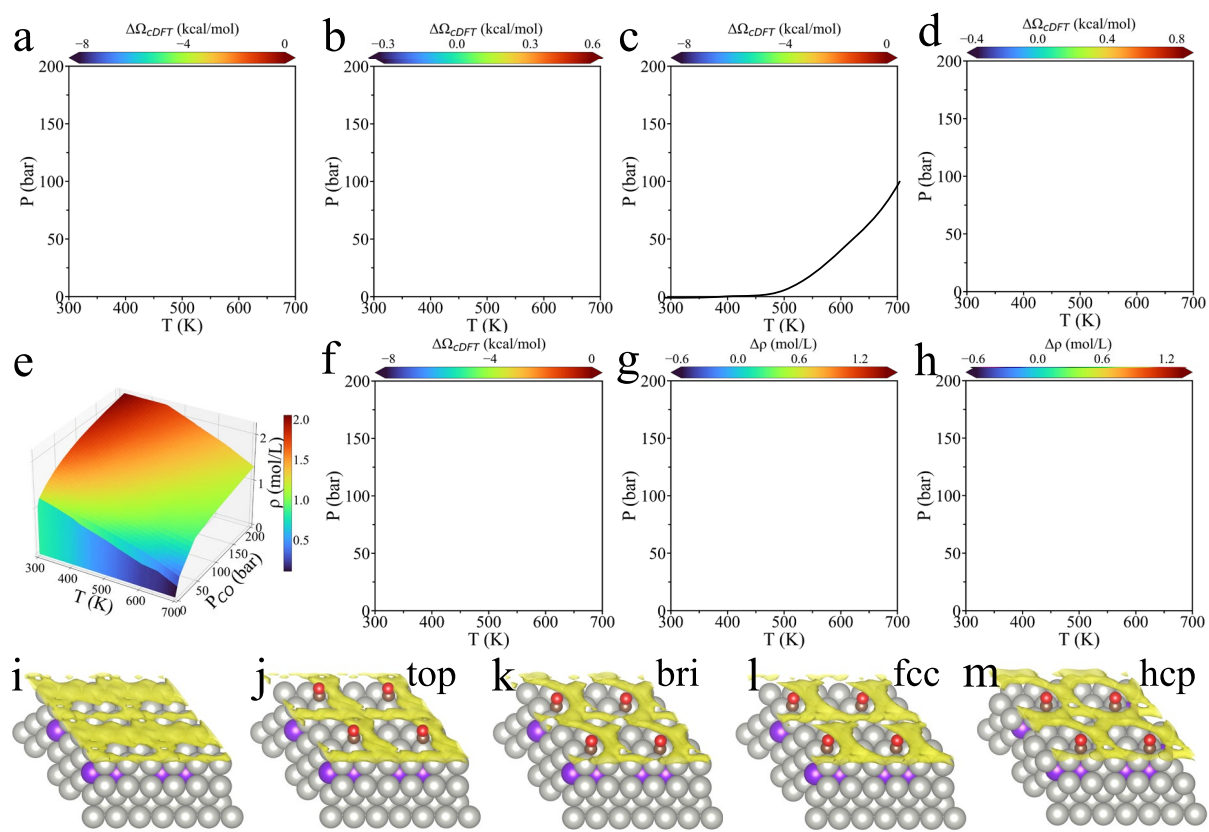

**Fig. 4** The penalty grand potential and the variation of CO density near several adsorption sites on the PdZn (111) surface under different temperatures and CO pressures. (a) The penalty grand potential ($\Omega_{cDFT-ad}$) when CO is bonded to the top site. The penalty grand potential difference ($\Delta\Omega_{cDFT}$) between (b) the bridge site and top site, (c) the fcc site and top site, and (d) the hcp site and top site. The black curve defines the pressure-temperature relationship at which $\Delta\Omega_{cDFT}$ reaches its maximum. (e) The surface density of CO when CO is bonded to the top site. Surface density differences ($\Delta\rho$) between (f) the bridge site and top site, (g) the fcc site and top site, and (h) the hcp site and top site. (i-m) Three-dimensional gas-phase CO density distribution when CO occupies different sites compared to the pristine surface at 500 K, $P_{CO}$ = 35 bar. The density iso-surface is set as 1 $e^{-6}$ mole/Å$^3$.

4e. A similar trend can be identified for the $\Omega_{cDFT-ad}$ values for other adsorption sites.

Additionally, we analyzed the environmental effects on different adsorption sites in terms of the difference in penalty grand potential, denoted as $\Delta\Omega_{cDFT}$. Fig. S4b, 4c, and 4d show the variation in the penalty grand potential for the bridge, fcc, and hcp sites relative to the top site. Compared to the top site, the other three sites exhibit a maximum difference at high temperature and low pressure. The black curve represents the pressure-temperature locus of maximum $\Delta\Omega_{cDFT}$. A positive value of $\Delta\Omega_{cDFT}$ indicates an increase in the penalty grand potential when CO moves from the top site to other sites, meaning that the free energy of gas-phase molecules physically adsorbed at these sites is higher than that at the top site. Therefore, at high temperature and low pressure, the environmental effect on the stability of surface CO follows the order: hcp > bridge > fcc > top. As temperature falls and pressure increases, $\Delta\Omega_{cDFT}$ gradually diminishes, and can become negative. At low temperature and high pressure, the environmental effect on the stability of gas-phase CO at the PdZn (111) surface changes to the order: hcp > top > bridge > fcc.

Fig. S4f, 4g, and 4h compares the surface density of gas molecules when CO bonds to the bridge, fcc, and hcp sites with the top site. We defined the surface density as the average density of CO molecules in the gas phase near the catalyst surface (within 6 Å), which more accurately reflects the gas distribution around the surface than the bulk density. At high temperature and low pressure, the variation of the surface density is negligible (the black curve indicates $\Delta\rho=0$). However, $\Delta\Omega_{cDFT}$ is at its maximum in this region, indicating that for the same surface density, the adsorption strength varies at different binding sites. As the temperature falls and pressure increases, the difference in the surface density of CO gradually increases, meaning that CO migration from the top site to the other sites is accompanied with an increase in the surface density of CO. This, in turn, suggests that the surface density of CO at the top site is the most concentrated. Combining this with the upper right side of the black curves in Fig. S4b, 4c, and 4d, it becomes evident that the increasing surface density leads to a gradual reduction in $\Delta\Omega_{cDFT}$, eventually causing a sign inversion. In other words, a higher surface density of CO results in a larger penalty grand potential. For the lower right portion of the black line, the gradual reduction of $\Delta\Omega_{cDFT}$ can be attributed to the reduction of the CO surface density.

The above analysis indicates that under temperature and pressure conditions near the black curves, the variation of surface density among different binding sites is relatively insignificant. In this regime, $\Delta\Omega_{cDFT}$ primarily depends on the inherent adsorption strength of CO on each site, following the order: hcp > bridge > fcc > top. As the temperature falls or pressure increases, the surface density of gas molecules at the top site rises more rapidly than at other sites, leading to a faster increase in $\Omega_{cDFT-ad}$. As a result, the difference in the penalty grand potential between the top site and other sites decreases and may even become negative. At low temperature and high pressure, $\Omega_{cDFT-ad}$ follows the order of hcp > top > bridge > fcc.

Additionally, we performed similar simulations for CO binding on the Pd (111) and Cu (111) surfaces. As shown in Fig. S1 for Pd (111), at high temperature and low pressure, the environmental correction follows the order of bridge ≈ fcc > hcp > top for the physical-sorption of gas-phase CO at different sites. At low temperature and high pressure, the order changes to bridge ≈ fcc = hcp > top. The surface density of CO follows the same trend as that adsorption on PdZn, remaining near zero under high-temperature, low-pressure conditions and gradually increasing as the temperature falls and pressure increases. For CO adsorption on the Cu (111) surface, the environmental effects follow the order: hcp > fcc > top > bridge, as shown in Fig. S2. The trends in the $\Delta\Omega_{cDFT}$ and surface density difference, as well as the underlying mechanisms, are similar to those observed for CO adsorption on the PdZn surface. It is generally believed that the bonding energy for the adsorption of CO at the different sites of the Cu (111) surface follows the sequence: top > fcc ≈ hcp > bridge.[35] This sequence reflects the order of surface bonding energy, whereas $\Omega_{cDFT-ad}$ is influenced by both the strength of the interaction between gas-phase molecules and the surface as well as the available contact volume at each site (i.e., the surface density of gas-phase molecules near the site). Therefore, the order of penalty grand potential differs from the order of bonding energy.

We also investigated the influence of $CO_2$ and $H_2$ partial pressures on CO adsorption at the four different sites on the catalyst surfaces mentioned above. The results are shown in Fig. SS4 and S5. Because $CO_2$ molecules adsorb more strongly on a metal surface than $H_2$ molecules, the penalty grand potential ($\Omega_{cDFT-ad}$) is primarily determined by the $CO_2$ partial pressure. Fig. S4 shows the differences in the adsorption grand potential ($\Omega_{ad}$) at each site. For PdZn (111) and Pd (111), the adsorption grand potential ($\Omega_{ad}$) exhibits the same trend as that predicted by KS-DFT calculations. For CO adsorption on the Cu (111) surface, the Gibbs adsorption free energy follows the order: fcc < hcp < bridge < top, as predicted by KS-DFT with the PBE functional. However, after correction with the penalty grand potential based on cDFT simulations, the binding sequence becomes bridge < fcc < hcp < top. This suggests that when the Cu (111) surface is in contact with $CO_2$ and $H_2$ molecules, CO prefer to adsorb at the bridge site even though the bonding energy is not as strong as other sites. This finding highlights a major difference between the conventional model of chemisorption and our grand potential method. In the presence of gas molecules in the reaction environment, chemisorption is essentially a substitution process. Gas molecules do not necessarily adsorb at the site with the lowest bonding energy. The adsorption site for a gas molecule depends on the difference in bonding energy ($\Delta E_{ad}$) at each site, as well as the change in the penalty grand potential ($\Delta\Omega_{cDFT-ad}$). The grand adsorption potential $\Omega_{ad}$ reflects both the bonding energy of the adsorbed species and the change in the local reaction environment, providing a more accurate prediction of adsorption sites.

*Preferential adsorption of CO on different surface sites*

By using the grand adsorption potential, we can analyze the environmental factors on the surface coverage of various chemical species in thermocatalytic reactions. In the following discussion, we use H and CO adsorption on the PdZn (111) surface as a case study.

Fig. 5a presents the surface phase diagram showing the coverages of H and CO, calculated according to the formation Gibbs energy ($G_{form}$) as explained in SI. Here the bonding energy data are obtained from the literature.[10] Both H and CO adsorb readily onto the PdZn (111) within the explored thermodynamic conditions (T=500 K, $P_{H2}$ 0.0001 bar – 10000 bar, $P_{CO2}$ 0.0001 bar – 1000 bar). For simplicity, we assume that the PdZn (111) surface is fully occupied by H and CO, so that the total surface coverage is normalized, i.e., $\theta_{CO} + \theta_H = 100$.

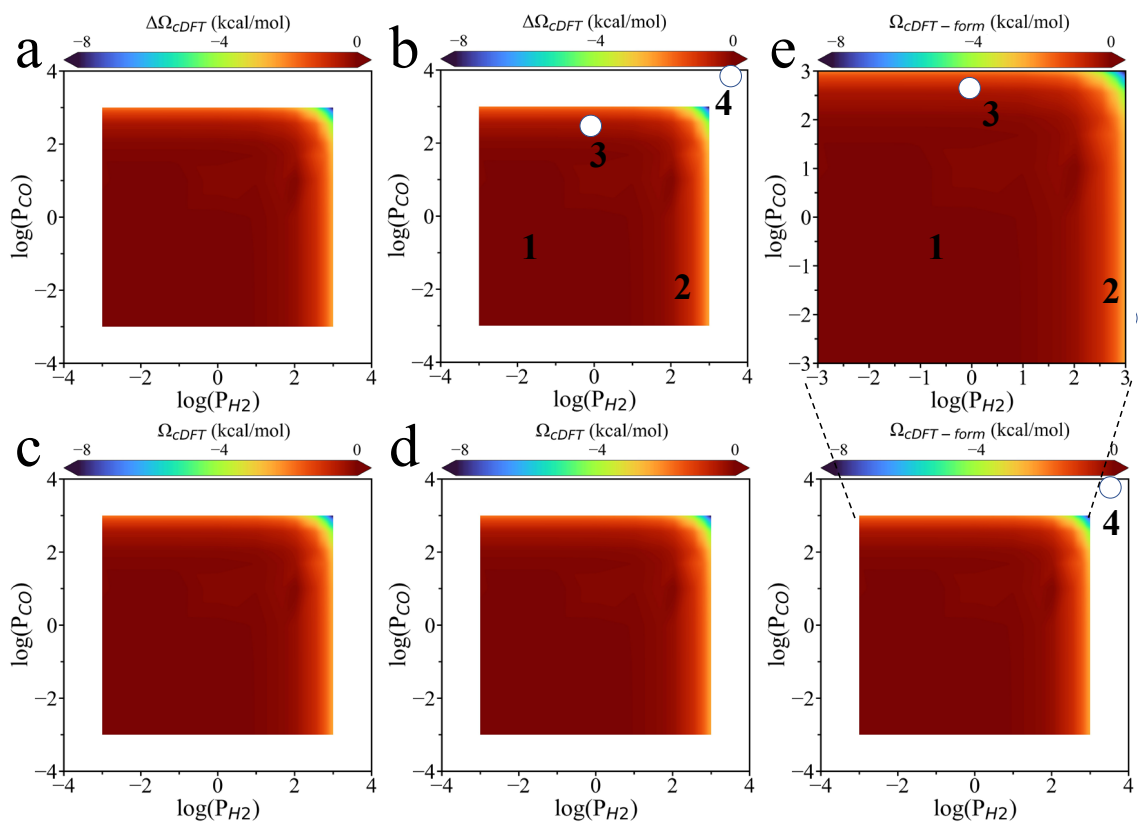

**Fig. 5** Surface coverage and free energy of formation under different $H_2$ and CO partial pressures. (a) Surface phase diagram predicted from the Gibbs energy of formation ($G_{form}$). (b) Surface phase diagram predicted from the grand potential of formation ($\Omega_{form}$). The surface is fully occupied by CO or H, i.e., $\theta_{CO} = 100 - \theta_H$, with $\theta_H$ shown in the color bar. Formation energy of PdZn_H075_CO025 ($\theta_H = 75$) calculated from the Gibbs energy (c) and from the grand potential (d). (e) Penalty formation grand potential ($\Omega_{cDFT-form}$) of PdZn_H075_CO025.

Fig. 5b displays revised surface phase diagram in terms of the surface coverages of these adsorbates after correction with cDFT simulations ($\Omega_{form}$). Although these surface phase diagrams have similar shapes, they predict different phase boundaries across specific temperature and pressure ranges, resulting in distinct stable surface phases. For example, after applying the grand potential correction, the H075_CO025 surface phase (i.e., $\theta_H = 75$, $\theta_{CO} = 25$) becomes less stable than the H100 phase ($\theta_H = 100$) at high $H_2$ partial pressures and low CO partial pressures (e.g., point 2 in Fig. 5b). At an $H_2$ partial pressure of 1 bar and a $CO_2$ partial pressure of approximately 800 bar (point 3), a portion of the H050_CO050 surface transitions to the H075_CO025 surface. Furthermore, the H025_CO075 surface phase expands in the high $CO_2$ partial pressure region, and 100% CO coverage becomes possible in the high $CO_2$ and high $H_2$ partial pressure region (point 4). A comparison of Fig. S5a and S5b reveals that the surface phase diagram obtained through the formation grand potential ($\Omega_{form}$) favors H coverage over CO at low pressures. The grand potential framework more accurately captures the sharp increase in CO coverage at ultra-high CO partial pressures, in better alignment with experimental observations.[17,36–38]

**Fig. S5c** and 5d show the formation energies of the H075_CO025 surface calculated through the Gibbs energy and the grand potential, respectively. Similar energy diagrams for other surface coverages can be found in Fig. S6. The ideal-gas correction for the partial pressure ($RT \ln P_i$) results in a linear relationship between the formation free energy ($G_{form}$) and the logarithm of the partial pressure for each adsorbate. However, the linear relationship does not hold for the formation grand potential ($\Omega_{form}$), highlighting the complexity of pressure effects due to intermolecular interactions. Fig. 5e illustrates the contribution of the environmental effects to the formation grand potential ($\Omega_{cDFT-form}$) for the H075_CO025 surface. Similar results for other surface coverages are presented in Fig. S6. At low $CO_2$ partial pressures, the formation free energy increases with increasing $H_2$ pressure. However, at high $CO_2$ partial pressures, the environmental correction is primarily influenced by the local density of $CO_2$. In this region, the competition between high $H_2$ and $CO_2$ partial pressures for adsorption sites results in a reduction of $\Omega_{cDFT-form}$. When the $H_2$ partial pressure is fixed, $\Omega_{cDFT-form}$ always increases with the $CO_2$ partial pressure, reducing its surface coverage.

To gain further insights into the stability of the adsorbates with the surface coverage at points 2, 3, and 4 in Fig. 5b, we analyzed the variation in the environmental effects on the formation grand potential $\Omega_{cDFT-form}$ across different coverage levels. As shown in Fig. 6a, when the $H_2$ partial pressure is fixed, a higher $CO_2$ partial pressure results in a greater the difference in $\Omega_{cDFT-form}$ between H075_CO025 and H100 surfaces. The high value of $\Omega_{cDFT-form}$ for H075_CO025 leads to its transition to H100 at point 2. Similarly, Fig. 6b shows that the H075_CO025 surface has a lower value of $\Omega_{cDFT-form}$ compared to H050_CO050, leading to the transition from H050_CO050 to H075_CO025 at point 3. Therefore, the surface phase transitions at point 2 and point 3 are driven by the penalty gran potential caused by the physical adsorption of gas-phase molecules. For point 4, the chemisorption of CO on the PdZn (111) surface increases the local densities of gas-phase molecules due to the high $CO_2$ and $H_2$ partial pressures. As shown in Fig. 5e, $\Omega_{cDFT-form}$ becomes a large negative value, below -400 kcal/mol, when the $CO_2$ and $H_2$ partial pressures approach 10,000 bar. Under these conditions, gas-phase molecules are strongly adsorbed at the catalyst surface, leading to a higher CO coverage.

Additionally, Fig. 6 shows that the difference in $\Omega_{cDFT-form}$ increases with increasing CO partial pressure for H100 transition to H075_CO025 as well as H075_CO025 transition to H050_CO050. However, $\Delta\Omega_{cDFT}$ essentially remains unchanged for the transitions from H050_CO050 to H025_CO075 and from H025_CO075 to CO100.

This is because for the transition from H100 to H050_CO050, the surface density of gas-phase species gradually falls as the CO coverage increases, as shown in Fig. S7. However, the surface density no longer changes for the transition from H050_CO050 to H025_CO075, indicating that at the H050_CO050 coverage, the surface is highly saturated and the penalty grand potential has reached its peak. As shown in Fig. S7, beyond this coverage, the surface density of gas molecule becomes similar to that of the bulk phase, indicating that CO has completely covered the PdZn surface. In other words, when 50% of the surface sites are occupied by CO, these CO molecules are sufficient to isolate the metal atoms from further interactions with the gas-phase environment. The surface saturation explains many experimental observations of CO poisoning effect on thermocatalytic reactions. For example, it aligns with the reduction of the dissociative probability of $D_2$ on nCO pre-covered Ru (0001) surface with increasing $\theta_{CO}$, with activity decreasing to nearly zero when $\theta_{CO}$ = 44.[39] Although a surface covered with H atoms can still interact with gas molecules in the reaction environment, it becomes effectively isolated from the gas phase once the CO coverage reaches 50%. The CO poisoning effect becomes particularly significant when the surface no longer sustains high concentrations of adsorbed species, leading to a drastic reduction in the catalytic activity.[16]

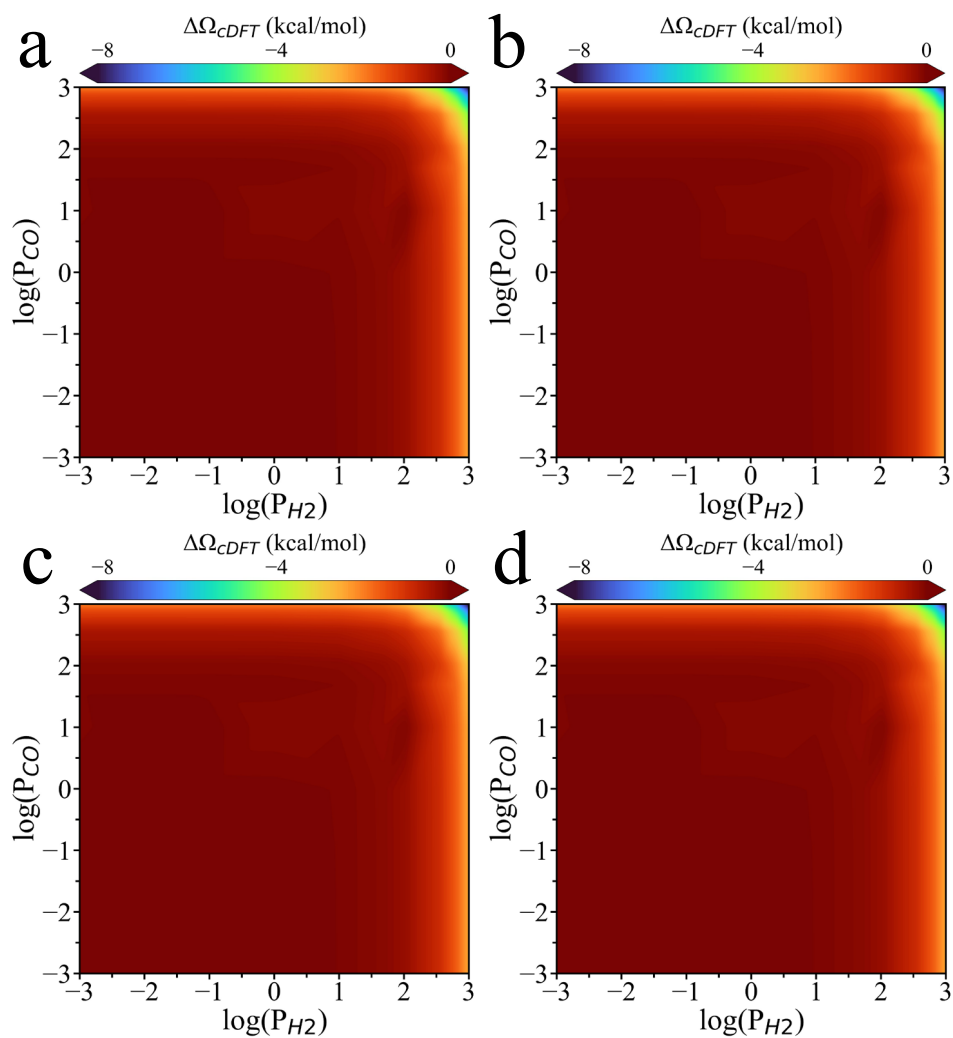

**Fig. 6** Difference in the penalty formation grand potential ($\Delta\Omega_{cDFT}$) between different surface coverages. (a) PdZn_H075_CO025 and PdZn_H100, (b) PdZn_H050_CO050 and PdZn_H075_CO025. (c) PdZn_H025_CO075 and PdZn_H050_CO050, (d) PdZn_CO100 and PdZn_H025_CO075.

## Conclusions

In this work we present a multiscale theoretical framework for calculating gas adsorption and surface coverage in thermocatalytic systems by combining KS-DFT calculations with classical DFT simulations. The adsorption grand potential ($\Omega_{ad}$) accounts for both the bonding energy and the variations in the grand potential of the gas phase using the grand canonical ensemble. To illustrate environmental effects on chemisorption, we have investigated the penalty grand potential ($\Omega_{cDFT-ad}$) for various species commonly involved in $CO_2$ hydrogenation on different catalytic surfaces. The theoretical results demonstrated the variation of $\Omega_{cDFT-ad}$ with the inherent adsorption strength of the catalyst surface, the surface density of gas molecules from the bulk phase, and the excluded-volume effects of the adsorbed species. Simulations of CO adsorption on the PdZn (111) revealed that the grand-potential correction of the adsorption energy can vary with the surface sites even under the same surface density of gas molecules from the bulk phase. Importantly, the grand-potential approach describes chemisorption as a substitution process, where the adsorption energy depends not only on bond formation but also on the grand potential penalty arising from the change of the reaction environment. The significant difference in the Gibbs-energy and grand-potential formations highlights the importance of considering the environmental effects in predicting the surface coverages of adsorption sites and reaction mechanisms, helping to better address the pressure gap in thermocatalytic studies.

## Author contributions

J.S. and J.W. designed the research, discussed the results, and wrote the paper. J.S. performed the calculations.

## Conflicts of interest

There are no conflicts to declare.

## Data availability

The authors have cited additional references within the Supporting Information.

## Acknowledgements

This research is made possible through financial support from the NSF-DFG Lead Agency Activity in Chemistry and Transport in Confined Spaces under Grant No. NSF 2234013. The authors thank Hong Zhang and Ping Liu for helpful discussions regarding the surface coverage on PdZn (111) surface.

## References

1  T. Goto, S. Ito, S. L. Shinde, R. Ishibiki, Y. Hikita, I. Matsuda, I. Hamada, H. Hosono and T. Kondo, *Commun. Chem.*, 2022, **5**, 1–10.
2  X. Liu, L. Sun and W.-Q. Deng, *J. Phys. Chem. C*, 2018, **122**, 8306–8314.
3  P. Wu and B. Yang, *ACS Catal.*, 2017, **7**, 7187–7195.
4  Y. Xi, T. Wang, J. Wang, J. Li and F. Li, *Catal. Sci. Technol.*, 2023, **13**, 6153–6164.
5  A. Cao, Z. Wang, H. Li, A. O. Elnabawy and J. K. Nørskov, *J. Catal.*, 2021, **400**, 325–331.
6  M. Schwarzer, D. Borodin, Y. Wang, J. Fingerhut, T. N. Kitsopoulos, D. J. Auerbach, H. Guo and A. M. Wodtke, *Science*, 2024, **386**, 511–516.
7  H.-X. Li, L.-Q.-Q. Yang, Z.-Y. Chi, Y.-L. Zhang, X.-G. Li, Y.-L. He, T. R. Reina and W.-D. Xiao, *Catal. Lett.*, 2022, **152**, 3110–3124.
8  M. Pineda and M. Stamatakis, *J. Chem. Phys.*, 2022, **156**, 120902.
9  B. Lacerda De Oliveira Campos, K. Herrera Delgado, S. Wild, F. Studt, S. Pitter and J. Sauer, *React. Chem. Eng.*, 2021, **6**, 868–887.
10  H. Zhang and P. Liu, *Chem Catal.*, 2025, **5**, 101156.
11  W. Xie, J. Xu, Y. Ding and P. Hu, *ACS Catal.*, 2021, **11**, 4094–4106.
12  Z. Wang, W. Xie, Y. Xu, Y. Han, J. Xu and P. Hu, *Catal. Sci. Technol.*, 2024, **14**, 5291–5303.
13  Gu K., Guo H. and Lin S., *Angew. Chem. Int. Ed.*, 2024, e202405371.
14  G. T. K. K. Gunasooriya and M. Saeys, *ACS Catal.*, 2018, **8**, 10225–10233.
15  S. Li, S. Rangarajan, J. Scaranto and M. Mavrikakis, *Surf. Sci.*, 2021, **709**, 121846.
16  M. Zhuo, A. Borgna and M. Saeys, *J. Catal.*, 2013, **297**, 217–226.
17  J. Fair and R. J. Madix, *J. Chem. Phys.*, 1980, **73**, 3480–3485.
18  V. Sumaria, L. Nguyen, F. F. Tao and P. Sautet, *ACS Catal.*, 2020, **10**, 9533–9544.
19  J. Sun, S. Jiang, Y. Zhao, H. Wang, D. Zhai, W. Deng and L. Sun, *Phys. Chem. Chem. Phys.*, 2022, **24**, 19938–19947.
20  Y. Tian, P. Hou, H. Zhang, Y. Xie, G. Chen, Q. Li, F. Du, A. Vojvodic, J. Wu and X. Meng, *Nat. Commun.*, 2024, **15**, 10099.
21  Y. Qin, C. Xia, T. Wu, J. Zhang, G. Gao, B. Y. Xia, M. L. Coote, S. Ding and Y. Su, *J. Am. Chem. Soc.*, 2024, **146**, 32539–32549.
22  J. Sun, R. Tu, Y. Xu, H. Yang, T. Yu, D. Zhai, X. Ci and W. Deng, *Nat. Commun.*, 2024, **15**, 6036.
23  J. Sun and J. Wu, *Chem. Sci.*, DOI:10.1039/D5SC00211G.
24  J. E. N. Swallow, E. S. Jones, A. R. Head, J. S. Gibson, R. B. David and M. W. Fraser, *J Am Chem Soc*.
25  K. Mori, T. Sano, H. Kobayashi and H. Yamashita, *J. Am. Chem. Soc.*, 2018, **140**, 8902–8909.
26  G. Yang, Y. Kuwahara, K. Mori, C. Louis and H. Yamashita, *Appl. Catal. B Environ.*, 2021, **283**, 119628.
27  X. Fang, Y. Men, F. Wu, Q. Zhao, R. Singh, P. Xiao, T. Du and P. A. Webley, *Int. J. Hydrog. Energy*, 2019, **44**, 21913–21925.
28  F. Abild-Pedersen, *Phys. Rev. Lett.*, DOI:10.1103/PhysRevLett.99.016105.
29  J. K. Norsko, *Rep. Prog. Phys.*, 1990, **53**, 1253.

30  G. Liu, A. J. Shih, H. Deng, K. Ojha, X. Chen, M. Luo, I. T. McCrum, M. T. M. Koper, J. Greeley and Z. Zeng, *Nature*, 2024, **626**, 1005–1010.
31  Y.-F. Shi, P.-L. Kang, C. Shang and Z.-P. Liu, *J. Am. Chem. Soc.*, 2022, **144**, 13401–13414.
32  Y. Kim, T. S. B. Trung, S. Yang, S. Kim and H. Lee, *ACS Catal.*, 2016, **6**, 1037–1044.
33  H. Wang, Y. Guo, G. Lu and P. Hu, *J. Chem. Phys.*, 2009, **130**, 224701.
34  M. Zhang, J. Ren and Y. Yu, *ACS Catal.*, 2020, **10**, 689–701.
35  Z. Chen, Z. Liu and X. Xu, *Nat. Commun.*, 2023, **14**, 936.
36  P. Zhao, Y. He, D.-B. Cao, X. Wen, H. Xiang, Y.-W. Li, J. Wang and H. Jiao, *Phys. Chem. Chem. Phys.*, 2015, **17**, 19446–19456.
37  T. Diemant, H. Rauscher, J. Bansmann and R. Jürgen Behm, *Phys. Chem. Chem. Phys.*, 2010, **12**, 9801–9810.
38  J. A. Davies and P. R. Norton, *Nucl. Instrum. Methods*, 1980, **168**, 611–615.
39  H. Ueta, I. M. N. Groot, M. A. Gleeson, S. Stolte, G. C. McBane, L. B. F. Juurlink and A. W. Kleyn, *ChemPhysChem*, 2008, **9**, 2372–2378.